\documentclass{article}

\usepackage[english]{babel}
\usepackage[a4paper, margin=1in, portrait]{geometry}

\usepackage{changepage}
\usepackage{amssymb}
\usepackage{amsmath}
\usepackage{graphicx}
\usepackage[colorlinks=true, allcolors=blue]{hyperref}

\usepackage[utf8]{inputenc}
\usepackage{graphicx} 
\usepackage{authblk} 

\title{RLEEGNet: Integrating Brain-Computer Interfaces with Adaptive AI for Intuitive Responsiveness and High-Accuracy Motor Imagery Classification}
\author{Sriram Nallani}
\author{Gautham Ramachandran}
\affil{Hypnos.ai: sriramnallani35@gmail.com, gauthamramachandran3@gmail.com}

\begin{document}
\maketitle
\setlength{\parindent}{0pt}
\setlength{\parskip}{\baselineskip}

\begin{abstract}
\noindent
Current approaches to prosthetic control are limited by their reliance on traditional methods, which lack real-time adaptability and intuitive responsiveness. These limitations are particularly pronounced in assistive technologies designed for individuals with diverse cognitive states and motor intentions. In this paper, we introduce a framework that leverages Reinforcement Learning (RL) with Deep Q-Networks (DQN) for classification tasks. Additionally, we present a preprocessing technique using the Common Spatial Pattern (CSP) for multiclass motor imagery (MI) classification in a One-Versus-The-Rest (OVR) manner. The subsequent 'csp\_space' transformation retains the temporal dimension of EEG signals, crucial for extracting discriminative features. The integration of DQN with a 1D-CNN-LSTM architecture optimizes the decision-making process in real-time, thereby enhancing the system's adaptability to the user's evolving needs and intentions. We elaborate on the data processing methods for two EEG motor imagery datasets. Our innovative model, RLEEGNet, incorporates a 1D-CNN-LSTM architecture as the Online Q-Network within the DQN, facilitating continuous adaptation and optimization of control strategies through feedback. This mechanism allows the system to learn optimal actions through trial and error, progressively improving its performance. RLEEGNet demonstrates high accuracy in classifying MI-EEG signals, achieving as high as 100\% accuracy in MI tasks across both the GigaScience (3-class) and BCI-IV-2a (4-class) datasets. These results highlight the potential of combining DQN with a 1D-CNN-LSTM architecture to significantly enhance the adaptability and responsiveness of BCI systems.
\end{abstract}

\section{Introduction}

\subsection{EEG Signals}
Electroencephalography (EEG) is a non-invasive method used to record electrical activity in the brain. When neurons communicate, they generate electrical impulses. EEG signals capture these impulses through sensors placed on the scalp. These signals are crucial for studying brain functions and have applications in medical diagnosis, neurology, and cognitive neuroscience. EEG is known for its high temporal resolution, despite having lower spatial resolution compared to other brain imaging techniques like fMRI [5, 9].

\subsection{EEG Head Caps}
An EEG headcap is a device used to hold electrodes in place on the scalp during EEG recordings. These head caps are designed with multiple electrode sites, conforming to a standardized layout known as the 10-20 system. This system ensures consistent placement of electrodes relative to specific areas of the brain. The electrodes in the headcap are usually made of conductive materials like silver chloride and are connected to an amplifier. The headcap makes it easier to set up the EEG by providing a standardized and more efficient way to place multiple electrodes, ensuring better consistency and reliability in the recordings.

\subsection{Motor Imagery}
Motor imagery refers to the mental simulation of movement without any actual movement or muscle activation. It involves imagining performing a task, such as moving a limb, which activates similar brain regions that are involved in the actual execution of the movement. This phenomenon makes motor imagery a valuable tool in neuroscientific research and applications like Brain-Computer Interfaces (BCIs), where it can be used to control external devices through brain signals alone.

\begin{table}[ht]
\centering
\begin{tabular}{l l l l}
\hline
Waveform & Frequency Range & Cognitive States & Typical Brain Regions \\
\hline
Delta & 0.5-4 Hz & Deep Sleep, Pathological States & Frontal, Parietal \\
Theta & 4-8 Hz & Light Sleep, Meditation & Temporal, Parietal \\
Alpha & 8-13 Hz & Relaxation, Calm Wakefulness & Posterior \\
Beta & 13-30 Hz & Active Thought, Alertness & Frontal, Central \\
Gamma & 30-100 Hz & High-level Cognition, Perception & Whole Brain \\
\hline
\end{tabular}
\caption{Frequency ranges and associated cognitive states for various EEG waveforms.}
\label{table:1}
\end{table}

\begin{figure}[ht]
\centering
\includegraphics[width=1\textwidth]{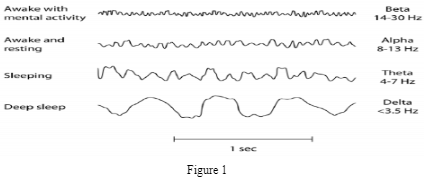}
\caption{EEG Waveforms \& Indication of Mental State}
\label{fig:example}
\end{figure}

\noindent {The important regions of the brain} - frontal-parietal, frontal, temporal, occipital, and central regions (denoted by FP, F, T, O, C, respectively in the channel nomenclature) - are involved in capturing EEG signals, the characteristics of which can serve as indicators of various cognitive states and neurological conditions, including seizures and other relevant neurodegenerative diseases [2,5]. For instance, individuals in deep sleep predominantly exhibit delta waves, while those in a meditative state show increased theta activity. During problem-solving tasks, beta wave activity is heightened, and gamma waves are associated with complex cognitive functions [9].

\noindent Table 1 shows the frequency ranges and associated cognitive states for various EEG waveforms. The asterisks indicate significant correlations with specific cognitive states or activities. The shaded band in the middle represents the typical frequency range for each waveform. One should note that upper and lower bounds differ across various sources.

Delta waves, ranging from 0.5-4 Hz, are most commonly observed during deep, non-REM sleep and are also indicative of certain pathological conditions. These waves are used for understanding unconscious bodily functions and are often used in sleep studies. Theta waves, with frequencies between 4-8 Hz, are associated with lighter stages of sleep, drowsiness, and meditative states. These waves are used for studying emotional responses and intuition and are often observed in the temporal and parietal regions. Alpha waves, in the 8-13 Hz range, are dominant when a person is awake but relaxed, often with eyes closed. These waves are used for understanding states of relaxation and are most prominent in the posterior regions of the brain. Beta waves, ranging from 13-30 Hz, are associated with active, analytical thought and are most commonly observed in the frontal and central regions. These waves are used for studying alertness and decision-making processes. Gamma waves, with frequencies above 30 Hz, are the fastest EEG waveforms and are associated with high-level cognitive functions. These waves are used for studying perception and consciousness [8, 9, 10, 19]. One can refer to information included in Figure 1 [1].

\subsection*{1.4 Existing Research}

\noindent Xiao and Fang (2021) detail the disadvantages of EEG signal processing such as low signal-to-noise ratio, and the fact that the several channels of the EEG signals do not have special meaning or function, which means that normal methods involve choosing a subset of the many channels, even though that may result in more noise or less distinctive feature extraction. The authors also propose a deep convolutional neural network architecture for highly accurate EEG signal classification on the BCI Competition Dataset along with physically acquired data. Their model attained an accuracy of 0.9324 for EEG signal recognition (precision: 0.9653; AUC: 0.9464) [28].

\noindent Saeidi et al. (2021) conducted a review of academic databases to gather relevant studies from 2000 onwards in order to measure the trends in ML and DL algorithms in terms of them being used in mental workload and motor imagery applications. They point out that 75\% of DL studies utilized convolutional neural networks in conjunction with many learning methods, and 36\% of ML studies used SVM to achieve a competitive accuracy. In the case of MI (motor imagery), the preferred and most frequently used algorithms included LDA (Linear Discriminant Analysis), CNN (Convolutional Neural Network), and SVM (Support Vector Machine) [21].

\noindent Venkatachalam et al. (2020) propose a Hybrid-KELM model which combines a Kernel Extreme Learning Machine (KELM) with Principal Component Analysis (PCA) and Fisher’s Linear Discriminant (FLD) for Motor Imagery classification of EEG data. The model used the BCI Competition III Dataset, attaining an accuracy of 96.54\%. The features listed in the work include entropy, phase, auto-regression, and spatial filters. They detail the two major classification methods: linear and nonlinear. The linear classification is done via LDA, while nonlinear classification is done with SVM and Neural Networks [25].

\noindent Altuwaijri and Muhammad (2022) propose the Multibranch EEGNet with Convolutional Block Attention Module (CBAM). When run upon the BCI-IV2a MI dataset and the high gamma dataset, the model attained the accuracies of 0.8285 and 0.9545, respectively. They outline how traditional deep learning lacks robust performance in EEG-MI classification. Also, they talk about how CNNs are becoming more popular in such applications, and that the DBNs (Deep Belief Networks), stacked autoencoders (SAEs), and RNNs are in vogue. The CNN allows for the spatio-temporal analysis and feature extraction, and provides a high accuracy on large datasets. This entails the fact that raw EEG signals can be inputted to a CNN, surpassing the statistical procedures such as kurtosis [2].

\noindent Mattioli et al. (2022) detail a highly accurate model made up of four one dimensional CNNs with interleaved BatchNormalization and Spatial Dropout layers, along with four Dense (Fully Connected - FC) layers with interleaved Dropout layers for a total of 10 layers. The researchers propose a 10-layer one-dimensional convolutional neural network (1D-CNN) for classifying five brain states (four MI classes plus a baseline class). This approach uses a limited number of EEG channels and a data augmentation algorithm. The paper also introduces a transfer learning method to extract critical features from a group EEG dataset and then customize the model for individual users by training its later layers with only 12 minutes of individual-related data, indicating the model’s adaptability to various users. When trained and tested upon the ‘'EEG Motor Movement/Imagery Dataset', the model achieved a 99.38\% accuracy at the group level [17].

\noindent Xie and Oniga (2023) review the use of CNNs for feature extraction in EEG signals. They note that while CNNs have been successful in other fields, their application in BCI, particularly in MI-EEG signal classification, is limited due to the need for large datasets, as well as difficulties in obtaining sufficient high-quality MI-EEG signals for training due to long-winded procedures and the noise / artifacts like eye and muscle movements. To address the issue of limited data, the study proposes a data augmentation method that enhances the volume of existing datasets. This method differs from traditional approaches by operating in both time and frequency domains, thus preserving the inherent structure of the data. The authors propose a combined time-frequency domain data enhancement method and a parallel input CNN. This model takes both raw EEG images and images transformed through continuous wavelet transform (CWT) as inputs. The approach aims to emphasize the main features of the original data while preserving other valuable features. The proposed algorithm reportedly achieves an average classification accuracy of over 97.61\% on the BCI Competition IV Dataset 2a, outperforming other algorithms that the authors examined [29].

\noindent Wiering et al. apply reinforcement learning (RL) to classification tasks, introducing a novel framework that models classification problems using Classification Markov Decision Processes and the MaxMin ACLA algorithm. Through extensive experiments with UCI datasets, the RL method, coupled with multi-layer perceptrons for function approximation, demonstrates competitive performance against traditional classifiers like multi-layer perceptrons and support vector machines. The findings suggest potential for RL in classification, highlighting the method's slight superiority over perceptrons and comparable efficacy to SVMs [26].

\section{Methodology}

\subsection{Data Preprocessing and Feature Extraction (GigaScience)}
The GigaScience dataset [7] involves EEG files collected from 25 healthy subjects over multiple recording sessions, specifically designed to capture a wide range of intuitive upper-extremity movement tasks. Each subject participated in three distinct recording sessions, which were conducted on different days, typically with a one-week interval between sessions. This setup aims to account for inter-session variability and to ensure the reliability and consistency of the data across time. Within each session, subjects were asked to perform 11 different movement tasks encompassing arm-reaching, hand-grasping, and wrist-twisting. These tasks were selected to cover a broad spectrum of upper-extremity movements and to generate a dataset that is comprehensive for BCI research.

Each movement task involved a specific number of trials to capture sufficient data for analysis. For example, arm-reaching tasks required 50 trials in each of the six directions, totaling 300 trials for arm-reaching alone per session. Across all subjects, the dataset encompasses a total of 82,500 trials, averaging approximately 3,300 trials per subject. This vast number of trials signifies the dataset's depth and its potential to provide substantial insights into brain activity related to different movements. The EEG data were recorded using a 60-channel setup according to the international 10-20 system. This extensive coverage ensures that the data capture a wide range of neural signals across the scalp, relevant to the motor and sensory processing of upper-extremity movements.

The EEG signals were sampled at 2500 Hz, with a 60 Hz notch filter applied to reduce the effect of electrical noise. This high sampling rate enables detailed time-series analysis of the brain's electrical activity during the performance of movement tasks. In addition to EEG, the dataset includes 7-channel electromyography (EMG) and 4-channel electrooculography (EOG) signals, making it a multimodal dataset. This inclusion allows researchers to explore the interactions between different types of physiological signals and their implications for BCI applications. The dataset's primary aim is to foster the development of intuitive BCIs that can interpret a user's intention based on natural movement thoughts or imagery. By providing a rich source of EEG, EMG, and EOG data related to various upper-extremity movements, the dataset supports research in areas such as neuro-rehabilitation, prosthetic control, and enhanced interaction with external devices.

We obtained 108 EEG Matlab files from this dataset, where 36 files corresponded to each of the three movements multigrasp, reaching, and wrist twisting. Our preprocessing involved selecting specific standard 10-20 system EEG channels C5, C3, C1, Cz, C2, C4, and C6, applying bandpass filtering (0.1-50 Hz) to reduce noise, and using the MNE Python library for data structuring. For feature extraction, we computed several statistical measures per channel and concatenated the Power Spectral Density for EEG channels C3 and C4 using the Welch method on the alpha and beta bands to measure ERD/ERS. The whole concatenated data yielded a shape of (370,). This data was normalized using MinMax normalization between -1 and 1 for consistency in training and testing phases. We performed a train-test-split with random shuffling and stratification to ensure similar numbers of samples from each of the 3 MI classes.

\subsection{Data Preprocessing and Feature Extraction (BCI-IV2a)}
The BCI Competition 2008 – Graz data set A (or Dataset 2a) [3] focuses on EEG data from 9 subjects participating in a cue-based BCI paradigm involving four different motor imagery tasks. These tasks include the imagination of movement of the left hand, right hand, both feet, and the tongue. The data set is structured into two sessions recorded on different days for each subject, with each session comprising 6 runs separated by short breaks. Each run consists of 48 trials, equally distributed among the four motor imagery classes, resulting in 288 trials per session, and 576 trials across the two sessions, and 5184 trials in the entire dataset.

At the beginning of each session, a recording of approximately 5 minutes was conducted to estimate the influence of EOG, divided into three blocks focusing on eyes open, eyes closed, and eye movements. EEG data were recorded using 22 Ag/AgCl electrodes following the 10-20 system, with a sampling rate of 250 Hz and a band-pass filter between 0.5 Hz and 100 Hz. Additionally, 3 monopolar EOG channels were recorded to facilitate artifact processing. Visual inspection by experts identified trials containing artifacts, marking them for potential exclusion or special processing in analyses.

Data sets are stored in the General Data Format (GDF), with one file per subject and session. Training data include class labels for all trials, while evaluation data, to be used for testing classifiers, do not. The competition emphasizes the importance of continuous classification output for each sample, including artifact-labeled trials. Evaluation is based on the time course of accuracy and the kappa coefficient, with the highest kappa value determining the winning algorithm.

We worked with each subject’s first session consisting of 287 trials, and excluded Subject 4, because both of that subject’s sessions had incomplete data (144 samples) in the respective GDF files A04T and A04E. The signals over all 22 channels were preprocessed with many iterations of ICA to remove artifacts. The Common Spatial Pattern was used in a One-vs.-The-Rest fashion, obtaining the most discriminative spatial filters for each of the four MI classes. The “csp\_space” transformation was used, preserving the number of time points per CSP spatial component, and 2D arrays of features, consisting of the same statistics stacked upon each other by component and the complete PSD estimated on the alpha and beta bands across all components. We altered the time windows, the number of CSP components, and the length of FFT of the Welch method for each subject to obtain the ideal number of features. We initially tested the CSP OVR method with an SVC classifier. The overall feature extraction process led to 2D feature arrays of shape (n\_csp\_components, n\_features\_per\_component) being created. These features were passed to the RLEEGNet model composed of DQN and 1D-CNN-LSTM.

\subsubsection{Welch’s Method}
The Welch method for estimating the power spectral density (PSD) of a signal involves several mathematical steps. Initially, the time series data, represented as \(x[n]\), is divided into overlapping segments. Each segment is then multiplied by a window function \(w[n]\). For a segment of length \(L\), the \(k\)-th segment, \(x_k[n]\), is given by \(x_k[n] = x[n + kD] \cdot w[n]\), where \(D\) is the overlap between segments and \(n\) ranges from 0 to \(L-1\).

The next step involves applying the Fourier transform to each windowed segment, resulting in a periodogram for each segment. The periodogram (f is frequency), \(P_k(f)\), is

\[
P_k(f) = \left| \frac{1}{L} \sum_{n=0}^{L-1} x_k[n] \cdot e^{-i 2\pi fn} \right|^2
\]

The Welch method's key feature is the averaging of these periodograms across all segments, which yields the PSD estimate, \(P_{xx}(f)\). When using median averaging, instead of taking the mean of these periodograms, the median is taken, as in the case of our feature extraction procedures. The equation then becomes:

\[
P_{xx}(f) = \text{median} \{ P_0(f), P_1(f), \ldots, P_{K-1}(f) \}
\]

where \(K\) is the total number of segments.

The PSD can be scaled in two ways: 'density' or 'spectrum'. In 'density' scaling, the PSD is normalized by the product of the segment length and the sampling frequency, resulting in units of power per frequency (e.g., \(V^2/Hz\)). This is represented as \(P_{xx}(f) = \frac{1}{L f_s} \cdot P_{xx}(f)\). In 'spectrum' scaling, the PSD is normalized by the square of the segment length, yielding units of power (e.g., \(V^2\)), calculated as \(P_{xx}(f) = \frac{1}{L^2} \cdot P_{xx}(f)\). The ‘density’ scaling option was used in our procedures.

Optionally, detrending can be applied to each segment before the Fourier transform. This process typically involves removing a linear or constant trend from the data. Finally, the method differentiates between real and complex signals. For real-valued signals, a one-sided spectrum is used, containing only positive frequencies [20].

\subsubsection{Kurtosis}
Kurtosis is a measure of the ``tailedness'' of the probability distribution of a real-valued random variable. The equation for the sample kurtosis is:
\[
\text{Kurtosis} = \frac{n(n+1)}{(n-1)(n-2)(n-3)} \sum_{i=1}^{n} \left( \frac{x_i - \bar{x}}{s} \right)^4 - \frac{3(n-1)^2}{(n-2)(n-3)}
\]
where \(n\) is the number of observations, \(x_i\) are the individual observations, \(\bar{x}\) is the mean of the observations, and \(s\) is the standard deviation [6, 15].

\subsubsection{Peak To Peak}
Peak to peak is a simple measure that calculates the difference between the maximum and minimum values in a dataset. It's given by:
\[
\text{Peak to Peak} = \max(x_i) - \min(x_i)
\]
Where \(x_i\) are the individual observations in the dataset [6, 15].

\subsubsection{Root Mean Square (RMS)}
The RMS is a measure of the magnitude of a set of numbers. It's particularly useful for calculating the average magnitude of varying values. 
The equation is:
\[
\text{RMS} = \sqrt{\frac{1}{n} \sum_{i=1}^{n} x_i^2}
\]
where \(x_i\) are the individual observations and \(n\) is the number of observations [6].

\subsubsection{Skewness}
Skewness measures the asymmetry of the probability distribution of a real-valued random variable. The equation for sample skewness is:
\[
\text{Skewness} = \frac{n}{(n-1)(n-2)} \sum_{i=1}^{n} \left( \frac{x_i - \bar{x}}{s} \right)^3
\]
Where \(n\) is the number of observations, \(x_i\) are the individual observations, \(\bar{x}\) is the mean of the observations, and \(s\) is the standard deviation [6, 15].

\subsubsection{Absolute Difference}
We also chose to use this feature. The absolute difference typically refers to the sum of the absolute differences between pairs of observations. It's calculated as:
\[
\text{Absolute Difference} = \sum_{i=1}^{n} |x_i - y_i|
\]
Where \(x_i\) and \(y_i\) are the individual observations in two datasets, and \(n\) is the number of paired observations.

\subsubsection{CSP and One-Vs.-Rest Application to Multiclass MI-EEG Classification}
The One-Versus-Rest (OVR) Common Spatial Pattern (CSP) approach is designed to tackle classification problems involving multiple classes, specifically \(N\) classes, by conducting \(N\) separate CSP analyses. In this methodology, each analysis uniquely identifies one class as the target class \(A\) and amalgamates the remaining classes into a single non-target class \(B\). This process results in the generation of \(N\) distinct sets of spatial filters, each set finely tuned to differentiate one specific class from all the others.

The operational framework of OVR CSP mirrors the foundational principles of traditional CSP but is distinctively applied in an iterative manner for each class against the collective rest. The procedure initiates with the calculation of the covariance matrix \(\Sigma_{\text{class } i}\) for the target class \(i\), followed by the computation of a combined covariance matrix \(\Sigma_{\text{rest}}\) representing the aggregate of all other classes. Subsequently, for each class \(i\), spatial filters \(w_i\) are determined through an optimization process that seeks to maximize the variance for class \(i\) while concurrently minimizing the variance for the combined rest of the classes, as expressed by the equation
\[
w_i^* = \arg\max_{w_i} \frac{w_i^T \Sigma_{\text{class } i} w_i}{w_i^T \Sigma_{\text{rest}} w_i}.
\]
Following the optimization, the spatial filters \(w_i^*\) are applied to the EEG signals to extract features that are most discriminative of each class \(i\) in comparison to the rest. The final step involves the classification of these extracted features.

By decomposing the multi-class problem into several binary classification tasks, the OVR CSP method effectively leverages the strengths of CSP in scenarios characterized by more than two classes of motor imagery tasks, thereby facilitating a more nuanced and precise classification process [27].

\subsection{1D-CNN-LSTM Architecture}
Our architecture integrates 1D Convolutional Neural Networks (1D-CNNs) with Long Short-Term Memory (LSTM) networks, creating a robust framework for time-series data analysis and classification. The model begins by reshaping the input data to align with convolutional processing requirements, followed by a batch normalization layer to stabilize the learning process, using the equation
\[
\hat{x}^{(k)} = \frac{x^{(k)} - \mu_\beta}{\sqrt{{\sigma_\beta}^2 + \epsilon}}
\]
where \(\mu_\beta\) and \({\sigma_\beta}^2\) denote the batch mean and variance, respectively, and \(\epsilon\) is a small constant to prevent division by zero.

\subsubsection{Conv1D Layers}
The core feature extraction is performed by two Conv1D layers [11, 17], with the first layer employing 64 filters of kernel size 7 to capture basic spatial patterns, and the second layer using 128 filters of kernel size 5 for more complex features. The Conv1D operation can be represented as
\[
y = f(W * x + b)
\]
where \(x\) is the input, \(W\) represents the convolutional filters, \(b\) is the bias, and \(f\) denotes the PReLU activation function, defined as
\[
f(y_i) = \max(0, y_i) + \alpha \min(0, y_i)
\]
for input \(y_i\) and a small coefficient \(\alpha\).

\subsubsection{PReLU (Parametric ReLU) Function}
PReLU (Parametric Rectified Linear Unit) enhances model learning by allowing small gradients when the unit is inactive, improving accuracy for deep networks by addressing the vanishing gradient problem. The function is also adaptive because it learns the coefficients of the negative part of its activation function during the training process, unlike traditional ReLU which has a fixed slope of zero for negative inputs. This adaptability allows PReLU to dynamically adjust its activation function to better fit the data, potentially leading to improved model performance on complex tasks.

\subsubsection{Dimensionality Reduction and Spatial Dropout}

Spatial feature processing is further refined with MaxPooling1D and AveragePooling1D layers, reducing dimensionality and emphasizing dominant features. SpatialDropout1D layers with a dropout rate of 0.1 are used to mitigate overfitting by randomly omitting feature maps.

Given an input tensor \(X\) of shape \([N, T, C]\), where \(N\) is batch size, \(T\) is the number of timesteps in the sequence, \(C\) is the number of channels/features and a specified pool size \(p\), the MaxPooling1D operation outputs a tensor \(Y\) with shape \([N, \frac{T}{p}, C]\) and each element \(y_{n,t',c}\) is computed as
\[
y_{n,t',c} = \max_{t=p(t'-1)+1}^{pt'} x_{n,t,c}.
\]
Here, \(x_{n,t,c}\) represents the value at the \(n\)-th sample in the batch, \(t\)-th timestep, and \(c\)-th channel of the input tensor \(X\). The operation applies the max function across each window of \(p\) timesteps for each channel \(C\), effectively capturing the most prominent signal within each window [17].

The AveragePooling1D operation, similar to MaxPooling1D, is used to reduce the dimensionality of the input sequence in time-series or sequence data processing. Unlike MaxPooling1D, which preserves the maximum value, AveragePooling1D calculates the average value over a specified window (pool size) along the temporal dimension. This operation smooths the input sequence, reducing its temporal resolution while retaining the average signal within each window. The underlying equation is:
\[
y_{n,t',c} = \frac{1}{p} \sum_{t=p(t'-1)+1}^{pt'} x_{n,t,c}.
\]
In this equation, \(x_{n,t,c}\) represents the value at the \(n\)-th sample in the batch, \(t\)-th timestep, and \(c\)-th channel of the input tensor \(X\). The operation computes the average across each window of \(p\) timesteps for each channel \(C\), providing a representation that reflects the general trend or average signal within each window [17].

\subsubsection{LSTM (Long Short-Term Memory Network)}

The LSTM layer, crucial for capturing temporal dependencies, incorporates 128 units with a \(\tanh\) activation function and is regularized using \(L1\) and \(L2\) penalties, formulated as \(L1L2(l1 = 0.01, l2 = 0.01)\), to balance learning complexity. The LSTM operates under the gates and state updates:
\begin{align*}
f_t &= \sigma(W_f \cdot [h_{t-1}, x_t] + b_f), \\
i_t &= \sigma(W_i \cdot [h_{t-1}, x_t] + b_i), \\
\tilde{C}_t &= \tanh(W_C \cdot [h_{t-1}, x_t] + b_C), \\
C_t &= f_t \ast C_{t-1} + i_t \ast \tilde{C}_t, \\
o_t &= \sigma(W_o \cdot [h_{t-1}, x_t] + b_o), \\
h_t &= o_t \ast \tanh(C_t),
\end{align*}
where \(f_t\), \(i_t\), \(o_t\) are the forget, input, and output gates, respectively, \(C_t\) is the cell state, \(h_t\) is the output vector, \(\sigma\) is the sigmoid activation function, and \(\tanh\) is the hyperbolic tangent activation function which normalizes its outputs between \(-1\) and \(1\), stabilizing the training [14, 16, 22].

\subsubsection{Global Max Pooling 1D}

Subsequent processing includes a GlobalMaxPooling1D (GMP1D) layer to distill essential temporal features across sequences. Given an input tensor \(X\) of shape \([N, T, C]\), where \(N\) is batch size, \(T\) is the number of timesteps in the sequence, \(C\) is the number of channels/features, the GMP1D layer outputs a tensor \(Y\) of shape \([N,C]\). Each element of this, \(y_{n,c}\), is
\[
y_{n,c} = \max_{t=1}^{T} x_{n,t,c}.
\]
Here, \(x_{n,t,c}\) represents the value at the \(n\)-th sample in the batch, \(t\)-th timestep, and \(c\)-th channel of the input tensor \(X\). The operation applies the max function across all timesteps \(T\) for each channel \(C\), effectively capturing the most prominent signal present at any timestep for each feature channel in the sequence.

This operation is beneficial for models dealing with variable-length input sequences where the exact position of important features within the sequence is less important than their presence. By reducing each feature channel to its maximum value over time, GMP1D allows subsequent layers of the model to focus on the most salient features from the sequence, improving the model's ability to make predictions based on the most critical information contained in the input data.

\subsubsection{Dense Layers}

Next are the dense layers with ReLU activation for complex pattern learning, incorporating \(L1/L2\) regularization to prevent overfitting. The dense layers' operations are defined as
\[
y = \text{ReLU}(W_h h + b_h)
\]
where \(W_h\) and \(b_h\) are the weights and bias of the dense layer, and \(h\) is the input from the previous layer.

The architecture concludes with an output layer, employing a linear activation function for regression or reinforcement learning outputs, and softmax for classification tasks, represented as
\[
y = \text{softmax}(W_o h + b_o)
\]
for the softmax activation, where \(W_o\) and \(b_o\) are the weights and bias of the output layer, respectively.

\subsection{Reinforcement Learning Aspects of Model}

We have developed a custom OpenAI Gym environment tailored for the reinforcement learning (RL) task involving classification of MI-EEG signals. The agent is a DQN which uses the Epsilon Greedy Policy with Exponential Epsilon Decay [23, 24]. This environment is characterized by a discrete action space consisting of the number of classes depending on the dataset, alongside an observation space that is one-dimensional for the GigaScience, and two-dimensional for BCI-IV2a. The design of this environment is particularly focused on RL applications, where each episode involves processing a specified dataset consisting of training data and corresponding labels.

In our environment, the core functionality is encapsulated within the 'step' function. This function represents a single time step within the RL task. When an action is input into this function, it processes this action and consequently returns several key pieces of information. Firstly, it provides the next observation, which is crucial for the RL agent to understand the current state of the environment [23].

Secondly, it calculates and returns a reward, which is based on the action taken. This reward is a critical component of reinforcement learning, as it guides the agent in learning which actions are beneficial towards achieving its goal. The cumulative reward is denoted in this manner:
\[
g_t = \sum_{k=0}^{\infty} \gamma^k r_{t+k}
\]
This equation illustrates the fundamental Q function, which represents the expected cumulative reward of taking action \(a\) in state \(s\).

The Bellman Equation for the Q function updates the Q value by considering the immediate reward \(r_t\) plus the discounted maximum future reward. In reinforcement learning, the quality of a state-action combination is evaluated using a function known as the Q function, denoted as \(Q^{\pi}(s, a)\). This function is defined as the expected cumulative reward
\[
\mathbb{E}^{\pi}[g_t | s_t = s, a_t = a]
\]
given that the agent is in state \(s\) and takes action \(a\).

According to the Bellman equation, the Q function can be further expressed as
\[
Q^{\pi}(s, a) = \mathbb{E}[r_t + \gamma \max_{a'} Q^{\pi}(s_{t+1}, a') | s_t = s, a_t = a]
\]
This formulation incorporates the immediate reward \(r_t\) plus the discounted future rewards, where \(\gamma\) is the discount factor (between 0 and 1), which influences whether the RL agent accumulates the beneficial short-term or long-term rewards [12, 13].

To maximize cumulative rewards, the agent aims to find the optimal Q function, denoted as \(Q^*\). The optimal policy, \(\pi^*\), under this optimal Q function, is considered the best strategy for action selection in a given state. This policy is 'greedy' because it always chooses the action that maximizes the current estimate of the Q function:
\[
\pi^*(a | s) = \begin{cases}
1 & \text{if } a = \arg\max_{a'} Q^*(s, a) \\
0 & \text{otherwise}
\end{cases}
\]

Finally, substituting this optimal policy into the Bellman equation, the optimal Q function \(Q^*\) is represented as
\[
Q^*(s, a) = \mathbb{E}[r_t + \gamma \max_{a'} Q^*(s_{t+1}, a') | s_t = s, a_t = a].
\]
This equation highlights that the value of a state-action pair under the optimal policy is the expected immediate reward plus the maximum expected discounted future rewards for the next state-action pair [13].

Thirdly, the step function returns a flag that indicates whether the episode has reached its conclusion. This flag is essential for the agent to understand when to terminate the current episode and potentially start a new one. Additionally, our environment includes a 'reset' function. This function is called at the beginning of each episode. Its primary role is to reset the environment to a standard initial state. This resetting process is crucial in reinforcement learning, as it ensures that each episode starts from a consistent state, allowing for the fair evaluation and comparison of different strategies and actions taken by the RL agent. Thus, this is reflective of the MDP - Markov Decision Process - a fundamental aspect of RL [26].

\subsubsection{DQN Online Q-Network as 1D-CNN-LSTM}

Utilizing a 1D-CNN-LSTM network as an Online Q-Network within a Deep Q-Network (DQN) framework is a novel approach to handling state representations, particularly for high-accuracy classification of Motor Imagery (MI) Electroencephalogram (EEG) signals. This hybrid architecture leverages the spatial feature extraction capabilities of 1D Convolutional Neural Networks (1D-CNNs) and the temporal sequence processing strengths of Long Short-Term Memory (LSTM) networks, making it exceptionally suited for interpreting the complex, time-sensitive patterns inherent in EEG data.

In the context of MI-EEG signal classification, the state representation is crucial for capturing the patterns within statistical and PSD features. Integrating the 1D-CNN-LSTM architecture as an Online Q-Network in a DQN setup involves using the network to predict the value of taking each possible action in a given state, based on the MI-EEG signal data. In this scenario, the actions could correspond to different classifications or decisions the system needs to make based on the MI-EEG signals. The state representation, facilitated by the 1D-CNN-LSTM network, provides a detailed understanding of the current MI-EEG signal, enabling the DQN to make informed decisions on the best action to take to maximize the expected reward, which in this case could be related to classification accuracy and relevant metrics.

The DQN framework allows the RLEEGNet to learn and adapt its strategy based on feedback, continuously improving its ability to classify MI-EEG signals accurately as more data is processed. The 1D-CNN-LSTM architecture (especially with with PReLU and GMP1D) is inherently designed to handle the variability and noise common in EEG data, making it robust against factors that could otherwise lead to misclassification. 

\subsubsection{Reward Structure for MI-EEG Signal Classification}

The concept of the reward structure is explored in [12], [13], [23], and [26]. We have designed the reward structure as follows:

\begin{itemize}
\item \textbf{Correct Classification Reward:} The model receives a positive reward for each correct classification of an MI-EEG signal. This reward encourages the model to make accurate predictions. The magnitude of the reward can be constant or vary based on the confidence of the classification.
\item \textbf{Incorrect Classification Penalty:} Conversely, the model incurs a penalty (negative or zero reward) for each incorrect classification. This penalty discourages the model from making the same mistake in the future and helps to fine-tune its decision-making process. Additionally, incorporating a zero reward for incorrect classifications within the reward structure of a DQN model for MI-EEG signal classification is a strategic decision that fosters a balanced learning environment. This approach promotes a distinction between not penalizing mistakes too harshly and rewarding correct predictions, facilitating a scenario where the model can safely explore various strategies without the deterrent effect of negative consequences for every error. It encourages the model to learn from its mistakes in a less punitive context, potentially leading to more robust learning outcomes.
\end{itemize}

The design of the reward structure is critical because it directly influences the learning and performance of the DQN model. A well-designed reward structure ensures that the model is properly incentivized to improve its classification accuracy over time. It must balance between encouraging correct classifications and penalizing incorrect ones, without discouraging the model from exploring potentially beneficial strategies that might initially result in errors. Through interaction with the environment (i.e., processing MI-EEG signals and receiving feedback based on the reward structure), the DQN model learns to adjust its internal parameters (including those of the embedded 1D-CNN-LSTM network) to maximize the cumulative reward. This process of trial and error, guided by the reward structure, enables the model to develop an effective strategy for classifying MI-EEG signals with high accuracy.

\section{Results of Evaluation on GigaScience and BCI-IV2a}

We present the results obtained via the RLEEGNet model for the GigaScience Multimodal EEG dataset and the BCI-IV2a dataset.

For the GigaScience EEG dataset, we experiment with 5 different reward combinations that inform the decision-making process of the model, where a positive reward indicates a correct classification and a negative (or zero) reward indicates wrong classification. The results are indicated in Table 2, where the means and standard deviations are given for each metric on the test set, 10 fold evaluation on the entire dataset (287 trials), and reward-based accuracy during its interaction with the RL-GYM environment based on the test set. An 80\%-20\% train-test split was used with stratification and random shuffling. The DQN training consisted of a first interval of 3000 steps for initial learning, and two separate intervals of 400 steps each to allow the model to exploit its learned optimal classification policy. These step numbers were selected to prevent the DQN from overfitting due to too many training steps.

For the GigaScience dataset, the DQN was compiled with Adam optimizer [10] with a learning rate of 0.0055 and a decay of 0.0001. For the BCI-IV2a dataset, the Adam optimizer had a learning rate of 0.0005 and decay of 0.001.

To test the CSP OVR method with ‘csp\_space’ transformation on the BCI-IV2a EEG dataset, the Tmin and Tmax values for the epochs were chosen along with the number of CSP components such that the One-versus-The-Rest (OVR) multiclass accuracy (computed by an SVC classifier) is maximized. Note: Anomalies in the data of Subject 4 resulted in an incomplete set of motor imagery EEG actions in the first session. No motor imagery actions were present in the second session except that of class 7 (left hand). Thus, the exclusion is because the available data for Subject 4 is not conducive to our examination of four class motor imagery classification.

Table 3 showcases the accuracy of an SVC classifier on four class motor imagery from the CSP OVR filters. The last row shows the average classification accuracy across the 8 subjects (1,2,3,5,6,7,8,9). It also shows the combination of number of CSP Components and epoch time window that yielded the maximum accuracy on a particular subject. The most informative time windows for CSP were selected as per the plots of STFT spectrograms that indicated pronounced brain activity.

Next, the CSP transformations were carried out on the epochs (retaining the same subject-specific epoching time windows and number of CSP components), but this time, the CSP-transformed EEG signals were projected onto the CSP Space, where the time domain is preserved in conjunction with the discriminative spatial components. The result of the “CSP Space” transformation yields a 3D array of shape (Trials, CSP Spatial Components, Time Points), where each trial has \(n\_\text{csp\_components}\) which each have the same number of time points as the original signal.

Similarly to the GigaScience dataset, the statistical features and the alpha and beta band’s power spectral density (via the Welch method) were computed such that they yielded a 2D shape of (CSP Spatial Components, Statistical + PSD features). These feature arrays were passed into the RLEEGNet model after a twofold z-score normalization with StandardScaler. The DQN training consisted of a first interval of 2500 steps for initial learning, and two separate intervals of 250 steps each to allow the model to exploit its learned optimal classification policy. These step numbers were selected to prevent the DQN from overfitting due to too many training steps. An 85\%-15\% train-test split was used with stratification and random shuffling. The reward structure used was (1,0) or (100.00\%, 0.00\%) in terms of (correct, incorrect). In Table 4, the means and standard deviations are given for each metric on the test set, 10 fold evaluation on the entire dataset (287 trials), and reward-based accuracy during its interaction with the RL-GYM environment based on the test set.

Table 5 showcases the confusion matrices for each subject, and table 6 shows the best combination of epoch time windows, number of time points, CSP components, and NFFT for Welch PSD method.

The training was remotely run on a workstation equipped with AMD Threadripper CPU (64 threads); GPU was inactive. The training time for the GigaScience dataset (80\% of 108 samples) was around 3-5 minutes, while the training time for the BCI-IV2a dataset (85\% of 287 samples) was around 10 minutes.
\clearpage
{Table 2: GigaScience Dataset Results (3 Class Classification)}
\begin{figure}[ht]
\includegraphics[width=\textwidth]{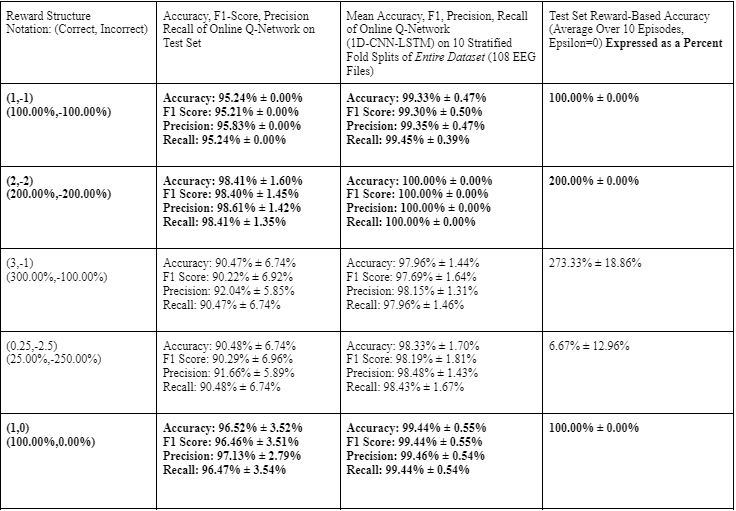}
\label{fig:fullwidth}
\end{figure}
\newpage
The variance in performance metrics across different reward structures in the GigaScience dataset can be attributed to the fundamental principles of reinforcement learning and how reward signals influence the learning process of RLEEGNet. In reinforcement learning, the reward structure plays a pivotal role in guiding the model towards desirable outcomes by reinforcing certain behaviors over others. This mechanism is particularly evident in the context of EEG signal classification, where the model learns to distinguish between three classes based on the feedback received from its actions. The excellent performance observed under the $(2, -2)$ reward structure, where the model achieved perfect scores, can be explained by the balanced and strong incentive for correct classifications coupled with a significant penalty for errors. This structure likely provided clear and consistent feedback that effectively guided the model's learning process, enabling it to optimize its parameters for maximum accuracy. The high reward for correct predictions and the equally high penalty for incorrect ones create a strong gradient for the model to follow, enhancing its ability to fine-tune its predictions towards more accurate classifications. Coupled with epsilon decay, the process of shifting from exploration to exploitation is all the more accelerated.

Conversely, the more challenging reward structures, such as $(3, -1)$ and $(0.25, -2.5)$, presented conditions that were not as conducive to optimal learning. For instance, the $(3, -1)$ structure, offering a high reward for correct classifications but a lower penalty for mistakes, might have led to a less stringent learning environment. This could allow the model to settle for suboptimal solutions, as the penalty for incorrect classifications was not severe enough to significantly deter the model from making errors. Similarly, the $(0.25, -2.5)$ structure, with a minimal reward for correct predictions and a disproportionately high penalty for errors, could create a highly cautious learning environment. This might encourage the model to prioritize avoiding mistakes over making correct classifications, potentially leading to underfitting or overly conservative predictions that do not fully capture the complexities of the EEG signal features. These observations underscore the delicate balance required in designing reward structures for reinforcement learning tasks.

The $(1, -1)$ reward structure, offering a balanced but moderate feedback mechanism, combined with epsilon decay, encourages initial exploration followed by a more focused exploitation of learned strategies. This approach yields a mean accuracy of $95.24\%$, indicating effective learning without overly discouraging risk-taking in the early stages. However, this is less than the accuracy of the $(1,0)$ and $(2,-2)$ reward structures. The $(1, 0)$ structure, offering a reward for correct classifications without penalizing incorrect ones, benefits significantly from the epsilon decay strategy. It supports extensive early exploration without the fear of negative consequences, leading to high accuracy and performance metrics as the model leverages its gathered knowledge in later stages of training.

The goal is to provide a reward signal that is neither too lenient nor too punitive but instead promotes a learning trajectory that aligns with the desired outcomes. The effectiveness of a reward structure is thus contingent upon its ability to calibrate the model's focus between exploring new strategies and exploiting known patterns, ultimately influencing the model's performance in complex classification tasks.
\clearpage
{Table 3: BCI Competition IV Dataset 2a Results (4 Class Classification)- OVR CSP}
\begin{figure}[ht]
\centering
\includegraphics[width=\linewidth]{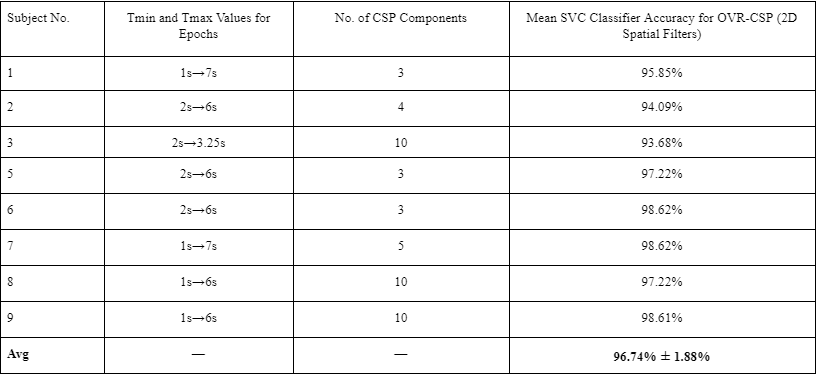}
\label{fig}
\end{figure}

The classifier accuracy on each subject of BCI-IV2a (except Subject 4) ranges from 93.68\% to 98.62\%, resulting in an average accuracy of 96.74\% ± 1.88\% across all subjects. The variation in temporal windows (Tmin to Tmax values for epochs) and the number of CSP components used per subject suggest a relationship between these parameters and classification accuracy. Subjects with broader temporal windows (1s to 7s) achieved high accuracies (up to 98.62\%), regardless of the number of CSP components used (3-5, 10). This suggests the choice of temporal window is crucial for capturing relevant EEG patterns.

Subject 3, with a narrower temporal window (2s to 3.25s) and more CSP components (10), had a slightly lower accuracy (93.68\%). This implies that while more CSP components can capture detailed spatial features, the effectiveness also heavily depends on the selected time window. For subjects with the temporal window ending at 6s but varying in start times and CSP components, accuracies were above 97\%, emphasizing the importance of optimizing both the temporal window and CSP components count. It shows that a well-chosen time window, coupled with an appropriate number of CSP components, is key to high classification performance. Subject 2, who had a temporal window of 2s to 6s and 4 CSP components, showed a lower accuracy (94.09\%), indicating that fine-tuning the temporal range and CSP feature dimensionality is critical.

*Refer to Table 6 for information on how the RLEEGNet requires at least 4-7 s of data besides the CSP components count and NFFT length for Welch PSD.

\begin{figure}[ht]
\begin{adjustwidth}{-5cm}{-4cm} 
\centering
Table 4: BCI Competition IV Dataset 2a (4 class classification)\\
\centering
[Reward Structure: +1 for correct classification, 0 for incorrect $\rightarrow$ (1,0)]
\includegraphics[width=35cm,height=25cm]{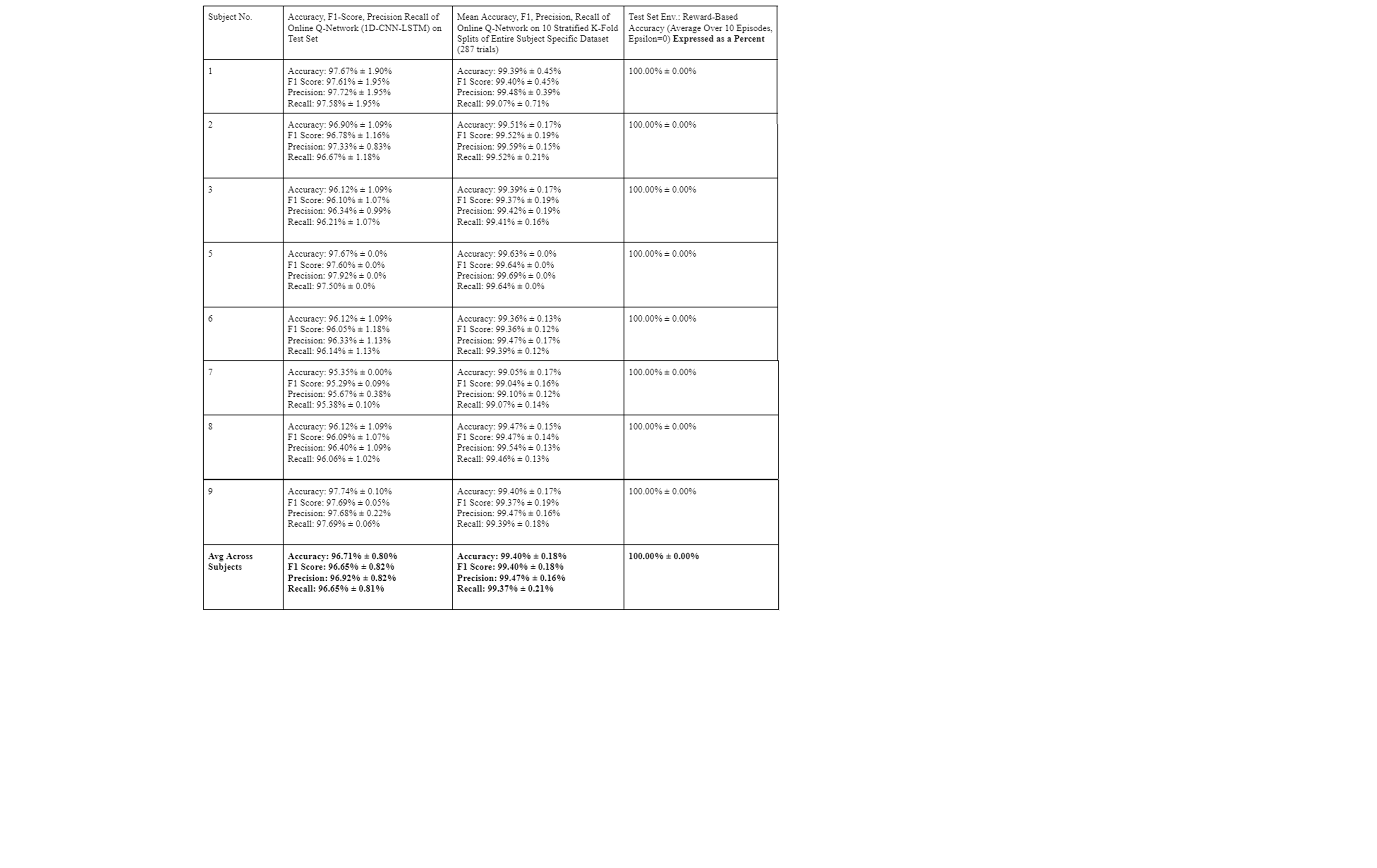}
\label{fig:example}
\end{adjustwidth}
\end{figure}
\clearpage
The average [intra-session] accuracy observed across individual subjects ranges from 95.35\% to 97.74\%. Such a variation not only attests to the model's capacity to classify EEG signals with a high degree of accuracy but also highlights the challenges inherent in achieving consistent performance across diverse datasets. The standard deviations associated with these metrics further illuminate the fluctuations in performance, reflecting the complexities of EEG data and the nuanced interactions between the model and this data. Analysis of the mean metrics obtained from K-fold validations, where accuracy rates are around 99\%, underscores the model's robustness and its ability to generalize across different segments of the dataset. Such consistently high mean accuracy rates, along with similarly high F1 scores, precision, and recall metrics, underscore the model's proficiency in maintaining performance consistency. This is a crucial attribute for brain-computer interface (BCI) applications, where accurate classification has significant implications.

The model achieves a uniform 100\% reward-based accuracy (maximum reward in the reward structure (1,0) ) across all subjects in a controlled testing environment with epsilon (exploration rate) set to 0. This reward, averaged over 10 episodes, underscores the model's capability to reach optimal classification outcomes under the subject-specific conditions. It suggests that the model can perform flawlessly within controlled parameters. The congruence of precision and recall metrics with accuracy and F1 scores, across both the individual tests and K-fold validations, indicates the model's balanced approach to classification. In EEG signal classification for BCI, where the system must differentiate between various signal categories, it is essential to maintain a balance between minimizing false positives and false negatives.    

Table 5: BCI Competition IV Dataset 2a Results (4 Class Classification)- Confusion Matrices\\
First Column: Subjects 1, 2, 3, 5; Second Column: Subjects 6, 7, 8, 9
\begin{figure}[ht]
\centering
\includegraphics[width=\linewidth]{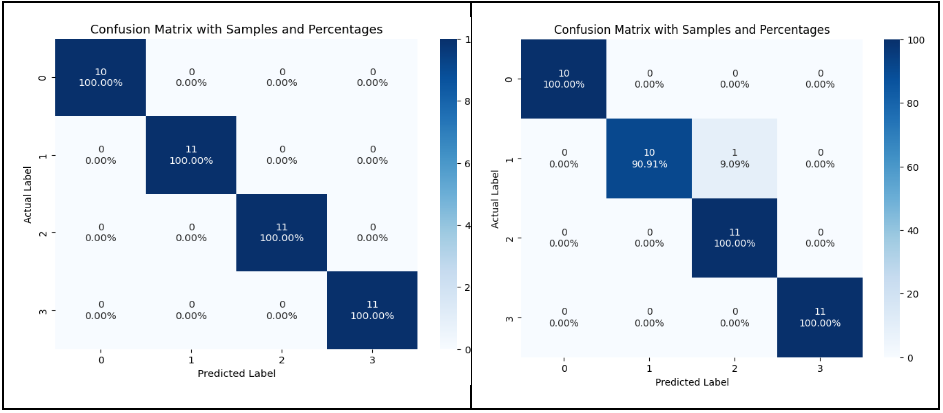}
\label{fig}
\end{figure}

\begin{figure}[ht]
\centering
\includegraphics[width=\linewidth]{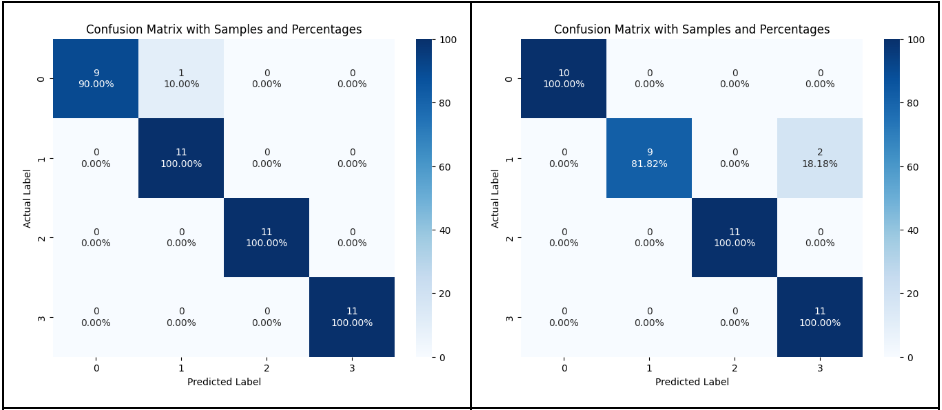}
\label{fig}
\end{figure}

\begin{figure}[ht]
\centering
\includegraphics[width=\linewidth]{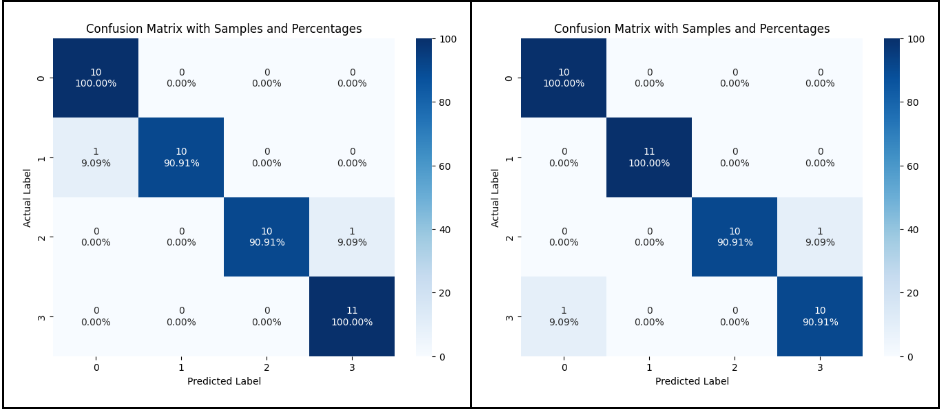}
\label{fig}
\end{figure}

\begin{figure}[ht]
\centering
\includegraphics[width=\linewidth]{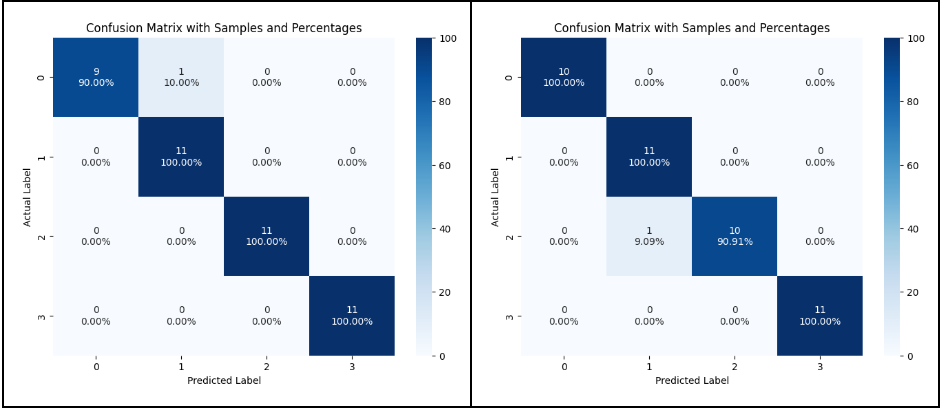}
\label{fig}
\end{figure}

\clearpage
Table 6: Subject-Specific CSP and Time Window Settings For RLEEGNet Model
\begin{figure}[ht]
\centering
\includegraphics[width=\linewidth]{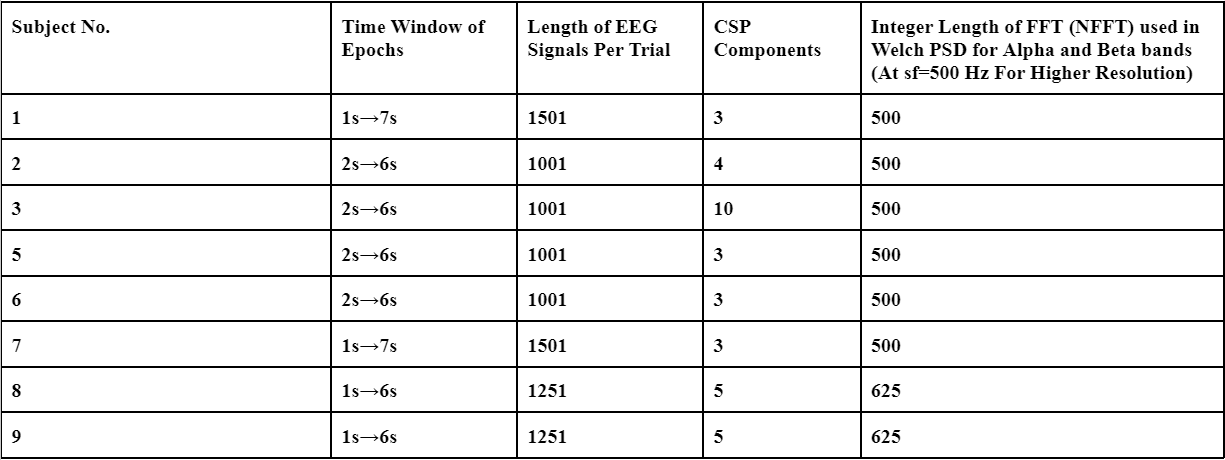}
\label{fig}
\end{figure}

The RLEEGNet demonstrated enhanced performance attributable to the longer signal durations. Specifically, for Subject 3, the application of a Support Vector Classifier (SVC) achieved a classification accuracy of 93.68\% within a temporal window spanning 2 to 3.25 seconds, along with 10 CSP components. Conversely, for Subjects 8 and 9, the SVC methodology required 10 CSP components to attain accuracies of 97.22\% and 98.61\%, respectively. However, the RLEEGNet approached these levels of accuracy with a reduced CSP components number. This phenomenon suggests that an optimal signal duration (4-6 seconds - difference between start and end times) combined with a moderate quantity of CSP components significantly contributes to the discriminative enhancement of the four Motor Imagery (MI) classes. The subsequent feature extraction conducted on the preserved time-domain signals, facilitated by the 'csp\_space', was sufficient to operationalize the RLEEGNet effectively.

\section{Conclusion}

Our research underscores the integration of DQN with 1D-CNN-LSTM for EEG signal classification within BCIs, highlighting the precision and adaptability of this approach. Notably, our novel CSP-OVR algorithm, which incorporates the 'csp\_space' transformation, significantly enhances spatial discriminability while retaining informative features in the time domain. This method, coupled with statistical and spectral feature extraction like Welch PSD, markedly boosts our model's accuracy in classifying MI-EEG signals within the challenging BCI-IV2a dataset, known for its artifacts and uncontrolled conditions [4]. The RLEEGNet attains state-of-the-art accuracy on both datasets, and on BCI-IV2a, the model surpasses the results in [2] by 13.86\% when considering the average [intra-session] accuracy across all subjects (96.71\%). The model attains the mean accuracy across all subjects between 95.91\% and 97.51\%, with the latter value being similar to the results in [29], but with less computational complexity in feature extractions and without any data augmentation. In the future, we envision our model using data augmentation, which can be a beneficial means to adapt RLEEGNet smoothly to classify new, continuously incoming signals through the BCI system.  

In-depth analysis reveals that optimizing accuracy depends on selecting appropriate temporal windows and CSP component counts. Moreover, the model's decision-making process benefits greatly from a well-structured reward system, indicating the intricate balance between reinforcement learning strategies and feature extraction methodologies in improving classification performance. Future work should aim to extend the model's applicability to a broader spectrum of EEG signals, and even specific, minor MI movements, enhancing its personalization and effectiveness in real-world BCI applications. Crucially, testing the model in real-time environments will be pivotal in validating its practical utility. Further exploration should also leverage RL to dynamically manage various methods of feature extraction and select optimal time windows and number of CSP components, ensuring the model's adaptability to varying signal characteristics and user-specific nuances. Additionally, the computational complexity can be better handled with specialized accelerators such as GPU and FPGA with efficient inference capabilities. In conclusion, our approach to classify MI-EEG signals promises to refine the understanding and application of BCIs, helpful in developing more sophisticated, user-centric solutions in neuroscience and AI integration. 

\section{References}
\begin{itemize}
    \item[] 1. Abhang, P., \& Gawali, B. W. (2015). Correlation of EEG images and speech signals for emotion analysis. \textit{ResearchGate}. Available from: https://www.researchgate.net/figure/EEG-waves-for-different-signals\_fig4\_281801676
    \item[] 2. Altuwaijri, G. A., \& Muhammad, G. (2022). Electroencephalogram-based motor imagery signals classification using a multi-branch convolutional neural network model with attention blocks. \textit{Bioengineering}, 9(7), 323. https://doi.org/10.3390/bioengineering9070323
    \item[] 3. Brunner, C., Leeb, R., Müller-Putz, G. R., Schlögl, A., \& Pfurtscheller, G. (n.d.). BCI competition 2008 – graz data set A. \textit{Institute for Knowledge Discovery, Graz University of Technology, Austria; Institute for Human-Computer Interfaces, Graz University of Technology, Austria.} Retrieved from https://lampx.tugraz.at/~bci/database/001-2014/description.pdf
    \item[] 4. Chowdhury, R. R., Muhammad, Y., \& Adeel, U. (2023). Enhancing cross-subject motor imagery classification in EEG-based brain–computer interfaces by using multi-branch CNN. \textit{Sensors}, 23(18), 7908. https://doi.org/10.3390/s23187908
    \item[] 5. Gosala, B., Kapgate, P. D., Jain, P., Chaurasia, R. N., \& Gupta, M. (2023). Wavelet transforms for feature engineering in EEG data processing: An application on schizophrenia. \textit{Biomedical Signal Processing and Control, 85}, Article 104811. https://doi.org/10.1016/j.bspc.2023.104811
    \item[] 6. Harpale, V., \& Bairagi, V. (2021). An adaptive method for feature selection and extraction for classification of epileptic EEG signals in significant states. \textit{Journal of King Saud University - Computer and Information Sciences, 33}(6), 668-676. https://doi.org/10.1016/j.jksuci.2018.04.014
    \item[] 7. Jeong, J.-H., Cho, J.-H., Shim, K.-H., Kwon, B.-H., Lee, B.-H., Lee, D.-Y., Lee, D.-H., \& Lee, S.-W. (2020). Multimodal signal dataset for 11 intuitive movement tasks from single upper extremity during multiple recording sessions. \textit{GigaScience, 9}(10), giaa098. \\https://doi.org/10.1093/gigascience/giaa098
    \item[] 8. Katmah, R., Al-Shargie, F., Tariq, U., Babiloni, F., Al-Mughairbi, F., \& Al-Nashash, H. (2021). A review on mental stress assessment methods using EEG signals. \textit{Sensors (Basel, Switzerland), 21}(15), 5043. https://doi.org/10.3390/s21155043
    \item[] 9. Kim, J. H., Rhee, J., Park, S., \& Lee, D. (2020). Exploratory study of brain waves and corresponding brain regions of professional gamers and amateur gamers. \textit{Scientific Reports, 10}(1), 1-11. https://doi.org/10.1038/s41598-020-64458-3
    \item[] 10. Kingma, D. P., \& Ba, J. (2017). Adam: A method for stochastic optimization. \textit{arXiv preprint arXiv:1412.6980}. Retrieved from https://arxiv.org/pdf/1412.6980.pdf
    \item[] 11. Kiranyaz, S., Avci, O., Abdeljaber, O., Ince, T., Gabbouj, M., \& Inman, D. J. (2021). 1D convolutional neural networks and applications: A survey. \textit{Mechanical Systems and Signal Processing, 151}, 107398. https://doi.org/10.1016/j.ymssp.2020.107398
    \item[] 12. Latif, S., Cuayahuitl, H., Pervez, F., Shamshad, F., Ali, H. S., \& Cambria, E. (2021). A survey on deep reinforcement learning for audio-based applications. arXiv. arXiv:2101.00240. Retrieved from https://arxiv.org/pdf/2101.00240.pdf
    \item[] 13. Lin, E., Chen, Q., \& Qi, X. (n.d.). Deep reinforcement learning for imbalanced classification. \textit{School of Computer Science and Engineering, South China University of Technology}. Retrieved from https://arxiv.org/pdf/1901.01379.pdf
    \item[] 14. Lu, Y., \& Salem, F. M. (2017). Simplified gating in long short-term memory (LSTM) recurrent neural networks. \textit{Michigan State University, East Lansing, Michigan}. Retrieved from https://arxiv.org/ftp/arxiv/papers/1701/1701.03441.pdf
    \item[] 15. Makkar, K., \& Bisen, A. (2023). EEG signal processing and feature extraction. \textit{International Journal for Modern Trends in Science and Technology, 9}(08), 45-50. \\https://doi.org/10.46501/IJMTST0908008
    \item[] 16. Mali, A., Ororbia, A. G., Kifer, D., \& Giles, C. L. (2021). Recognizing and verifying mathematical equations using multiplicative differential neural units. In \textit{Proceedings of the Thirty-Fifth AAAI Conference on Artificial Intelligence (AAAI-21)}. \\Retrieved from https://ojs.aaai.org/index.php/AAAI/article/view/16634/16441
    \item[] 17. Mattioli, F., Porcaro, C., \& Baldassarre, G. (2021). A 1D CNN for high accuracy classification and transfer learning in motor imagery EEG-based brain-computer interface. \textit{Journal of Neural Engineering, 18}(6), 066053.
    \item[] 18. Medina, V., \& Başar-Eroğlu, C. (2009). Event-related potentials and event-related oscillations during identity and facial emotional processing in schizophrenia. \textit{International Journal of Psychophysiology, 71}(1), 84–90. https://doi.org/10.1016/j.ijpsycho.2008.07.010
    \item[] 19. Nayak, C. S., \& Anilkumar, A. C. (2023). EEG normal waveforms. In \textit{StatPearls [Internet]}. Treasure Island (FL): StatPearls Publishing. \\Retrieved from https://www.ncbi.nlm.nih.gov/books/NBK539805/
    \item[] 20. Rashida, M., \& Habib, M. A. (2023). Quantitative EEG features and machine learning classifiers for eye-blink artifact detection: A comparative study. \textit{Neuroscience Informatics, 3}, 100115. https://doi.org/10.1016/j.neuri.2022.100115
    \item[] 21. Saeidi, M., Karwowski, W., Farahani, F. V., Fiok, K., Taiar, R., Hancock, P. A., \& Al-Juaid, A. (2021). Neural decoding of EEG signals with machine learning: A systematic review.\textit{ Brain Sciences, 11}(11), 1525. https://doi.org/10.3390/brainsci11111525
    \item[] 22. Sak, H., Senior, A., \& Beaufays, F. (n.d.). Long short-term memory recurrent neural network architectures for large scale acoustic modeling. [Google, USA]. Retrieved from \\https://static.googleusercontent.com/media/research.google.com/en//pubs/archive/43905.pdf
    \item[] 23. Sutton, R. S., \& Barto, A. G. (2018). \textit{Reinforcement learning: An introduction} (Second edition). The MIT Press. \\Retrieved from https://www.andrew.cmu.edu/course/10-703/textbook/BartoSutton.pdf
    \item[] 24. Tokic, M. (2010). Adaptive e-greedy exploration in reinforcement learning based on value differences. In \textit{Proceedings of the 7th German Conference on Multiagent System Technologies} (pp. 203-215). Springer, Berlin, Heidelberg. Retrieved from \\https://www.researchgate.net/publication/221562563\_Adaptive\_e-Greedy\_Exploration\\\_in\_Reinforcement\_Learning\_Based\_on\_Value\_Differences
    \item[] 25. Venkatachalam, K., Devipriya, A., Maniraj, J., Sivaram, M., Ambikapathy, A., \& Amiri, I. S. (2020). A novel method of motor imagery classification using EEG signal. \textit{Artificial Intelligence in Medicine, 103}, 101787. https://doi.org/10.1016/j.artmed.2019.101787
    \item[] 26. Wiering, M. A., van Hasselt, H., Pietersma, A.-D., \& Schomaker, L. (n.d.). Reinforcement learning algorithms for solving classification problems. \textit{University of Groningen}. Retrieved from https://www.ai.rug.nl/~mwiering/GROUP/ARTICLES/rl\_classification.pdf
    \item[] 27. Wu, W., Gao, X., \& Gao, S. (2005). One-versus-the-rest(OVR) algorithm: An extension of common spatial patterns (CSP) algorithm to multi-class case. In \textit{Conference proceedings: ... Annual International Conference of the IEEE Engineering in Medicine and Biology Society}. IEEE Engineering in Medicine and Biology Society. Annual Conference, 2005, 2387–2390. https://doi.org/10.1109/IEMBS.2005.1616947
    \item[] 28. Xiao, X., \& Fang, Y. (2021). Motor imagery EEG signal recognition using deep convolution neural network. \textit{Frontiers in Neuroscience, 15}. https://doi.org/10.3389/fnins.2021.655599
    \item[] 29. Xie, Y., \& Oniga, S. (2023). Classification of motor imagery EEG signals based on data augmentation and convolutional neural networks. \textit{Sensors, 23}, 1932. https://doi.org/10.3390/s23041932 
\end{itemize}

\end{document}